\documentclass[%
reprint,
amsmath,amssymb,
aps,
prb,
onecloumn,
]{revtex4-2}

\usepackage{amsmath,amssymb}
\usepackage{graphicx}
\usepackage{dcolumn}
\usepackage{bm}
\usepackage{xcolor}
\usepackage[colorlinks,linkcolor=blue,citecolor=blue,urlcolor=blue]{hyperref}

\renewcommand\thesection{\arabic{section}}
\usepackage{titlesec}

\begin{document}
	
	\preprint{APS/123-QED}
	
	\title{Purity-dependent Lorenz number, electron hydrodynamics and electron-phonon coupling in WTe$_2$}
	
	\author{Wei Xie$^{1}$, Feng Yang$^{1}$, Liangcai Xu$^{1}$, Xiaokang Li$^{1}$, Zengwei Zhu$^{1,*}$ and Kamran Behnia$^{2,*}$}
	
	\affiliation{$^1$Wuhan National High Magnetic Field Center and School of Physics, Huazhong University of Science and Technology,  Wuhan,  430074, China\\
		$^2$Laboratoire de Physique et d'\'{E}tude des Mat\'{e}riaux (ESPCI - CNRS - Sorbonne Universit\'{e}) PSL University, Paris, 75005, France
	}
	\date{\today}
	
	\begin{abstract}
		We present a study of electrical and thermal transport in  Weyl semimetal WTe$_2$ down to 0.3 K. The Wiedemann–Franz law holds below 2 K and a downward deviation starts above. The deviation is more pronounced in cleaner samples, as expected in the hydrodynamic picture of electronic transport, where a fraction of electron-electron collisions conserve momentum. Phonons are the dominant heat carriers and their mean-free-path do not display a Knudsen minimum. This is presumably a consequence of weak anharmonicity, as indicated by the temperature dependence of the specific heat. Frequent momentum exchange between phonons and electrons leads to quantum oscillations of the phononic thermal conductivity. Bloch-Gr\"uneisen picture of electron-phonon scattering breaks down at low temperature when Umklapp ph-ph collisions cease to be a sink for electronic flow of momentum. Comparison with semi-metallic Sb  shows that normal ph-ph collisions are amplified by anharmonicity. In both semimetals, at cryogenic temperature, e-ph collisions degrade the phononic flow of energy but not the electronic flow of momentum.

		\textbf{thermal conductivity, Wiedemann-Franz law, electron hydrodynamics, electron-phonon coupling}

		\textbf{PACS number(s):}51.20.$+$d, 63.20.Kr, 75.47.De

	\end{abstract}

	\maketitle

	\section{Introduction}
	Semi-metallic WTe$_2$  attracted much attention, first because of its large magnetoresistance \cite{ref1,ref2,ref3,ref4}, and then following its identification as a type-II Weyl semimetal \cite{ref5}. It has a carrier density as low as $n=p \simeq 6.8 \times 10^{19}$ cm$^{-3}$ \cite{ref2}. This implies that several hundred primitive cells share a single mobile electron (as well as a mobile hole). Its large orbital magnetoresistance is due to the high mobility of carriers whose long  wavelength attenuates scattering by point-like defects. Thanks to compensation\cite{ref44}, magnetoresistance does not saturate in the high-field limit. Such features are also detectable in other semi-metals such as antimony  ($n=p= 5.5 \times 10^{19}$ cm$^{-3}$)\cite{ref6,ref7} and  in WP$_2$ ($n=p= 1.5 \times 10^{21}$ cm$^{-3}$)\cite{ref8,ref9}. 
	
	Recently, thermal transport in both WP$_2$\cite{ref8,ref9} and in Sb\cite{ref6,ref10} has been studied in order to detect signatures of hydrodynamics.
	These are expected when a significant portion of collisions between particles conserve momentum instead of relaxing it. This idea was first put forward,  decades ago by Gurzhi \cite{ref11}, who proposed the possibility of viscous flow of electrons in metals and phonons in insulators. A renewal of interest in this topic has led to the experimental scrutiny of thermal transport by electrons \cite{ref6,ref8,ref12} and by phonons \cite{ref10,ref13,ref14,ref15}, as well as a number of theoretical studies\cite{ref16,ref17,ref18,ref19}. 
	
	Here, we present a study of electrical ($\sigma$) and thermal ($\kappa$) conductivities in bulk WTe$_2$ single crystals with different residual resistivities from 100 K down to 0.3 K.  We quantify the Lorenz ratio $L=\frac{\kappa}{T\sigma}$ and find that it exceeds the Sommerfeld value of  $L_0 =\frac{\pi^2}{3}\frac{k_B^2}{e^2}$. This  means that, in our range of investigation, thermal conductivity by phonons dominates heat transport. Using a magnetic field, we can separate the electronic ($\kappa_e$) and the phononic ($\kappa_{ph}$) components of the thermal conductivity. We find that $\frac{\kappa_e}{T\sigma}\simeq L_0$ at 2 K, thus, the Wiedemann–Franz (WF) law holds in the zero temperature limit, when inelastic scattering is absent. As the temperature increases, $\frac{\kappa_e}{T\sigma}$ begins to fall below $L_0$.  Thermal resistivity, defined as $(\kappa/T)^{-1}$ follows $T^2$ with a prefactor larger than that of the $T^2$ prefactor of electrical resistivity, as previously reported in other metals \cite{ref6,ref9,ref20,ref21,ref22,ref23}. By comparing samples with different residual resistivities, we find that the deviation from the WF law increases with the increase of the mean-free-path, as expected in the hydrodynamic picture of heat transport \cite{ref16}. 
	
	The phononic thermal conductivity, $\kappa_{ph}$, of WTe$_2$, displays quantum oscillations,  indicating significant electron-phonon coupling, providing an explanation for the absence of ballistic phonon transport down to 0.3 K. In contrast to antimony\cite{ref6,ref10}, there is no local Knudsen minimum in the temperature dependence of phonon mean-free-path, indicating the absence of phonon hydrodynamics. Furthermore, unlike antimony, the phonon specific heat displays an asymptotic Debye $T^3$ temperature dependence, indicating  weak normal ph-ph scattering  in WTe$_2$. Nevertheless, like in antimony, below a threshold temperature, the Bloch-Gr\"unesien  picture of electron-phonon resistivity is suddenly interrupted. Below this temperature, the electronic flow do not loose momentum because of e-ph collisions.

	\section{Results and discussion}
\begin{figure}[t]
		\centering
		\includegraphics[width=8cm]{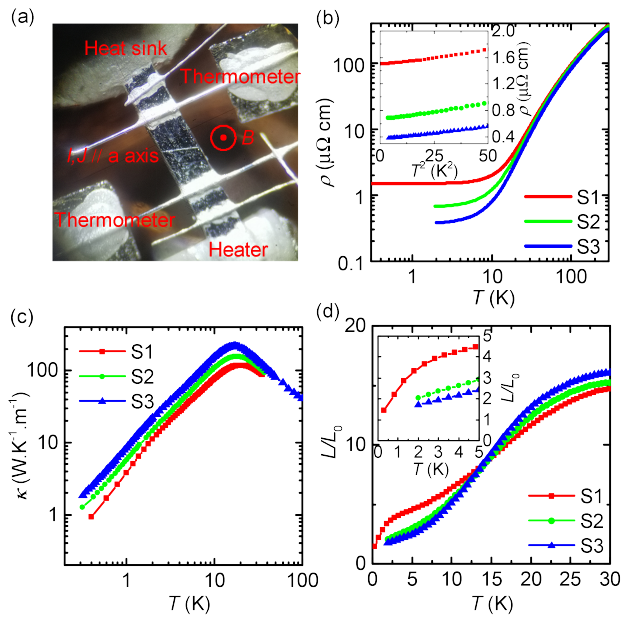}
		\caption{\textbf{Electrical and thermal transport in absence of magnetic field.} (a) A photograph of the sample, the contacts for measuring local temperature and electric field and the thermometers. (b) Temperature dependence of the electrical resistivity of three WTe$_2$ samples. The inset shows the low temperature data as a function of  $T^2$. (c) Temperature dependence of the thermal conductivity of three WTe$_2$ samples. (d) The  $L/L_0$ ratio as a function of temperature. The inset is a zoom on the low temperature data displaying the recovery of the WF law in the zero temperature limit.}
		\label{fig1}
	\end{figure}

	As shown in Figure \ref{fig1}a, we used a standard one-heater-two-thermometers method to measure thermal conductivity (for details see the supporting information). Electrical and heat currents were applied along the $a$ axis and the magnetic field was oriented perpendicular to them.  Figure \ref{fig1}b presents the temperature dependence of electrical resistivity in three different WTe$_2$ samples with RRR ratio ($\rho$(300 K)/$\rho$(2 K)) of 250, 540 and 840 respectively. The carrier mean-free-path $\ell_{0}$, estimated by the Drude formula is in the range of 3.2-13.2 $\mu$m, smaller than the size of our samples. The inset shows the same data as a function of $T^2$. One can see that resistivity varies quadratically with temperature. The temperature dependence of thermal conductivity ($\kappa$) of the same samples are shown in figure \ref{fig1}c. $\kappa$  peaks in all three around 18 K. Below this maximum, $\kappa$ quickly decreases. The temperature dependence of $L/L_0$ is presented in Figure \ref{fig1}d. Here  $L$ is the  experimentally measured Lorenz number and $L_0$ is the Sommerfeld value set by fundamental constants. Within our temperature range of measurements, $L/L_0$ $>$ 1, but  when the temperature is close to zero, it tends towards unity in conformity with the WF law. 
	
	The scrutiny of the  $L/L_0$ is instructive. In WP$_2$, another Weyl semi-metal, $L/L_0$ becomes less than unity at cryogenic temperatures  \cite{ref8,ref9}, and tends towards in the zero-temperature limit \cite{ref9}. This means that the phononic contribution to the thermal transport is negligible in WP$_2$, which has approximately 20 times more carriers than WTe$_2$. On the other hand, in graphite \cite{ref15} (which has 20 times less carriers) and in Sb \cite{ref10} (which has almost the same carrier density), the $L/L_0$ ratio exceeds unity when the sample is warmed. These are the cases where the phononic component of thermal conductivity accounts for a significant part of the total thermal conductivity. In addition to the phonon contribution, ambipolar contribution can also enhance the Lorenz ratio\cite{ref24}, but only when carriers are warmed above their degeneracy temperature, which is not the case here. 
	
	In order to separate  phononic and electronic components of the total thermal conductivity, we can exploit magnetic field and the fact that it influences these two components in a very different manner. This procedure  has been previously employed by several authors in a variety of semi-metals \cite{ref6,ref25,ref26,ref27}. 
	
\begin{figure}[h]
		\centering
		\includegraphics[width=8cm]{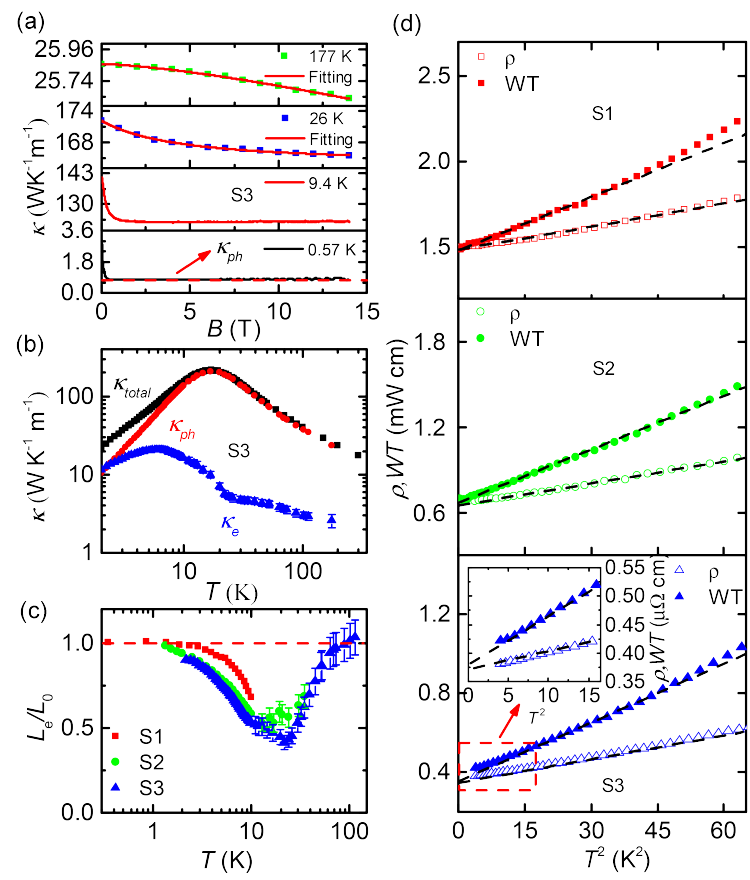}
		\caption{\textbf{The departure from the WF law and the $T^2$ resistivities.} (a) Magnetic field dependence of the thermal conductivity of sample S3 at several different temperatures. The solid lines in red show the fitting of formula (1), and the electron and phonon thermal conductivity can be separated by the fitting. The dashed line represents that the phonon thermal conductivity does not change with the magnetic field after the electron thermal conductivity is suppressed.(b) The temperature dependence of electronic, phononic and total thermal conductivity of sample S3. Error bars are due to the fitting errors.(c) The temperature dependence of $L_e/L_0$, $L_e$=$\kappa_e\rho$/T is the electronic Lorenz number, and $L_0$ is the Sommerfeld constant. (d) Thermal and electrical resistivities plotted as functions of $T^2$, the solid symbols represent thermal resistivity and the hollow symbols represent electrical resistivity. The black dashed line is a linear fitting to the data, and the slopes of the fitting to the solid and hollow symbols are the T-square prefactors of the thermal and electrical resistivities, respectively. The inset is a zoom on the low temperature data.}
		\label{fig2}
	\end{figure}

	Figure \ref{fig2}a shows the  field dependence of thermal conductivity at different temperatures. One can see that at low temperature, thermal conductivity decreases rapidly with magnetic field and then becomes flat. The electronic thermal conductivity is rapidly suppressed under the magnetic field (reflecting the very large electrical magnetoresistance of the system). In contrast to  this,  the phonon thermal conductivity remains almost unaffected. At 10 K, a 3 T magnetic field is high enough to suppress the amplitude of the electronic thermal conductivity by many orders of magnitude. Therefore, we can obtain electron thermal conductivity ($\kappa_{e}$) by subtracting the thermal conductivity at 3 T (which is equivalent to $\kappa_{ph}$) from the total thermal conductivity at zero-magnetic-field ($\kappa_{\rm{{total}}} (B=0)$). 
	
	As the temperature rises, the electron (and hole) mobility decreases, the thermal magnetoresistance of electrons becomes less drastic and the separation of the phononic and electronic components becomes less straightforward. In this temperature range the field dependence of thermal conductivity can be fitted by this empirical expression \cite{ref10}: 
	
	\begin{equation}
		\kappa_{total}=\kappa_{ph}+{\frac{T}{\alpha+\beta{B^\gamma}}}
		\label{1}
	\end{equation}
	
	The first term on the right side of the equation corresponds to the field-independent phononic component, $\kappa_{ph}$. The second term corresponds to the field-dependent electronic component. The functional form of this second term mimics the field dependence of a magnetoresistance with a power law field dependence. Among the three fitting parameters, only $\alpha$ and $\beta$ vary with temperature. $\gamma \approx 1.45$ remains constant in our range of investigation. Similar empirical expressions have been widely used in various materials \cite{ref10,ref27,ref28,ref29,ref30}, in most of these materials, the electronic component is a small fraction of the total thermal conductivity.
	Figure \ref{fig2}b shows $\kappa_{e}$ and $\kappa_{ph}$  extracted for sample S3 with this procedure. $\kappa_{ph}$ is larger than $\kappa_{e}$ above 3 K and dominates the thermal conductivity, consistent with the ratio $L/L_0$ exceeding unity with warming (Figure \ref{fig1}d). 
	
	Having separated the two components of thermal conductivity, let us now consider the electronic Lorenz number: $L_e=\kappa_e\rho/T$. The evolution of $L_e/L_0$ with temperature in different samples is presented in Figure \ref{fig2}c, at temperatures below 2 K and above 100 K, $L_e/L_0 \approx 1$, that is the WF law is recovered. However, around 20 K, $L_e/L_0 \ll 1$. Comparing the three samples, one sees that, in the cleanest sample (S3) $L_e/L_0$ shows the strongest deviation  from unity. 
	
	The finite temperature downward deviation from the WF law is traditionally attributed to the presence of inelastic small-angle scattering. In presence of inelastic scattering, thermal and electrical transport are affected in different ways, referred to as “horizontal” and “vertical” processes \cite{ref9,ref31}. The thermal resistivity is affected by the two processes while the electrical resistivity is only affected by the horizontal process. As a consequence, $L_e/L_0$ will be degraded with warming. In the hydrodynamic picture,  momentum-conserving collisions are also considered. In this case, momentum-conserving electron–electron collisions, which do not affect the electric current flow would still influence the heat flow, pulling down the $L_e/L_0$ ratio.
	
	As in the case of Sb \cite{ref6}, a way to distinguish between  these two scenarios, is to study samples with different levels of purity. In the hydrodynamic picture, the deviation from the WF law becomes more pronounced with the relative abundance of momentum-conserving electron–electron collisions compared to electron-boundary or electron-defect collisions. On the other hand, the relative weight of e-e small-angle scattering, which depends on Fermi surface geometry and the screening length \cite{ref31} is not expected to change with residual resistivity. Therefore, our observation that the attenuation of $L/L_0$ amplifies with sample purity is in agreement with the expectations of the hydrodynamic picture \cite{ref16} and in agreement with the detection of a Poiseuille flow of electrons in this system \cite{ref32}.
	
	Inelastic e-e scattering can degrade the momentum flow by two known mechanisms\cite{ref33,ref34}. The first is through Umklapp collisions, which are rare in WTe$_2$.  Umklapp events cannot occur when $4k_F^{max} < G$, the width of the Brillouin zone is $G=2\pi/c=4.5 nm^{-1}$ ($c$=1.4 nm is the lattice parameter). Considering the mild anisotropy of the Fermi surface \cite{ref2}, we get $4k_a =2.2$ nm$^{-1}, 4k_b =2.9$ nm$^{-1}, 4k_c =5.0$ nm$^{-1}$. Only the latter is slightly larger than $G$. Therefore,  Umklapp collisions are possible but rare. The second mechanism, proposed by Baber \cite{ref35} requires multiple Fermi surfaces and can operate in WTe$_2$.

	\begin{figure*}[t]
		\centering
		\includegraphics[width=17cm]{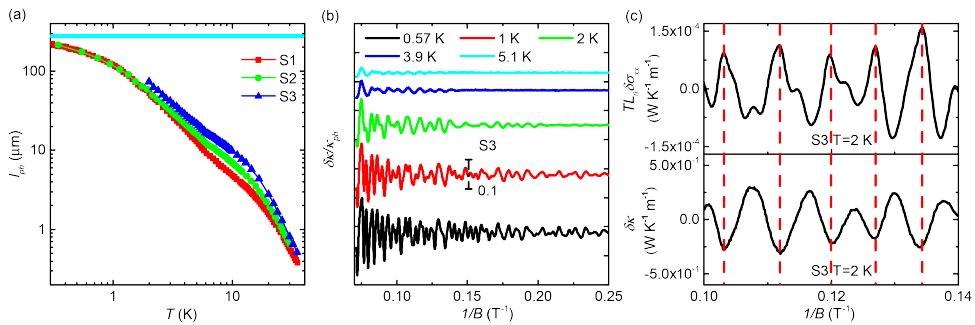}
		\caption{\textbf{The phonon mean-free-path and quantum oscillations.} (a) The phononic mean-free-path as a function of temperature. The mean-free-path is much shorter than the effective thickness of the samples shown by the light blue horizontal line. This implies that  phonons are not ballistic, but scattered by electrons. (b) The normalized amplitude of the oscillatory  thermal conductivity,  $\delta\kappa/\kappa_{ph}$  as a function of $\textit{1}/B$ for various temperatures. Data for different temperature are shifted for clarity. The scale bar corresponds to a relative amplitude of 10$\%$. (c) Quantum oscillations of thermal conductivity compared to oscillations of electrical conductivity in comparable units,  $TL_0\delta\sigma_{xx}$. Red dashed lines correspond to the maximum and minimum of the two sets of oscillations, which are out of phase.}
		\label{fig3}
	\end{figure*}
	
	Figure \ref{fig2}d shows the temperature dependence of electrical resistivity, $\rho$ and thermal resistivity,  $WT=(\frac{\kappa}{L_0T})^{-1}$, for the three samples. The horizontal axis is the square of temperature. One can see that in all cases, $\rho$ follows $\rho = \rho_0 + AT^2$  and $WT$ follows $WT = (WT)_0 +BT^2$, $\rho_0$ and  $WT_0$ are the residual resistivity associated with impurities. Their amplitude is locked to each other by the WF law.
	$A$ and  $B$ are  the prefactor of the $T^2$ resistivity. As seen in the figure, while the intercepts are identical, the slopes are different implying that $B>A$. Similar behavior has been observed in many metals \cite{ref22}, such as Sb\cite{ref6}, WP$_2$\cite{ref9}, W\cite{ref20}, CeRhIn$_5$\cite{ref21}, SrTi$_{1-x}$Nb$_x$O$_3$ \cite{ref23} and UPt$_3$\cite{ref36}. In all these cases, $B>A$, compatible with both scenarios of the Lorenz number. However, to the best of our knowledge, only in antimony and in the present study, an evolution of $B/A$ ratio with the variation of residual resistivity has been sought and found. As discussed in the supporting information, this variation is weaker in WTe$_2$ than in Sb. 
	
	Let us now consider the phonon thermal conductivity $\kappa_{ph}$. As  seen in Figure \ref{fig2}b, it dominates thermal conductivity in almost all our range of investigation. Figure \ref{fig3}a shows the temperature dependence of the extracted phonon mean-free-path  using $\ell_{ph}=\frac{3\kappa_{ph}}{\left\langle v_s\right\rangle C_{ph}}$ (See supporting information for details). $\ell_{ph}$ increases with decreasing temperature, It tends to become comparable to the sample size well below 0.3 K, which is our lowest temperature of investigation. What impedes phonon to become ballistic at $T \approx$ 4 K? Phonon-phonon scattering, can decay heat current only if collisions are Umklapp. At 4 K, the typical wave-vector of an acoustic phonon ($\frac{k_BT}{\hbar v_s}$) is too short compared to the width of the Brillouin zone to allow Umklapp scattering. The wavelength of phonons is also too long to allow scattering by point defects. In crystalline insulators  phonons become ballistic in this temperature \cite{ref14,ref37}. Phonons are either scattered by extended disorder or by mobile electrons.
	
	Examination of the quantum oscillations of electrical and thermal transport in WTe$_2$ provides a clue.  The oscillatory part of the thermal conductivity, obtained after subtracting a  smooth background, is shown in Figure \ref{fig3}b. (See the supplement for the Fourier transform and the discussion of main frequencies). Figure \ref{fig3}c compares this oscillatory part of the thermal conductivity, $\delta\kappa$ with Shubnikov-de Haas oscillations of the electrical conductivity normalized with the WF law. One can see that the amplitude of oscillations in $\delta\kappa$ is 3 orders of magnitude larger than in $TL_0\delta\sigma_{xx}$ (the order of magnitude of the electronic component). The maximum and minimum of the $TL_0\delta\sigma_{xx}$ and $\delta\kappa$ are out of phase. Similar features were reported in TaAs \cite{ref38}, NbP \cite{ref39}, TaAs$_2$ and NbAs$_2$ \cite{ref40}, where  $\delta\kappa$ was found to be 2 orders of magnitude larger than $TL_0\delta\sigma_{xx}$. In Sb \cite{ref10}, a 5 orders of magnitude discrepancy was found. In the latter case, the oscillations were attributed to phonons and their strong coupling with electrons. This is backed by the fact that oscillations are out of phase. Each time electrical conductivity has a peak due to a peak in the density of states caused by the evacuation of a Landau level, the phonon thermal conductivity shows a minimum. The case of WTe$_2$ is similar. Mobile electrons are therefore the main reason for the short mean-free-path of phonons at cryogenic temperatures, seen in Figure \ref{fig3}a (See the supporting information for a discussion of the Dingle mobility and its orders of magnitude discrepancy with transport mobility).
	
	\begin{figure*}[ht]
		\centering
		\includegraphics[width=17cm]{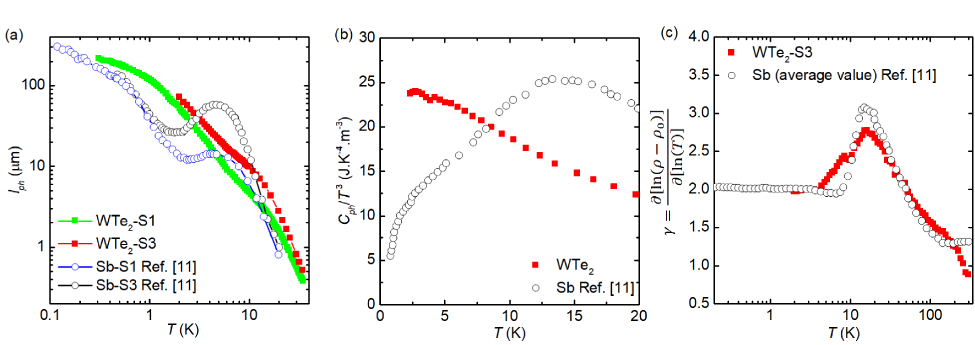}
		\caption{\textbf{Comparison with antimony.} (a) The temperature dependence of phonon mean-free-path, $\ell_{ph}$ in WTe$_2$ and in Sb \cite{ref10}. In both cases, the phonon mean-free-path remains much smaller than the sample size due to scattering by electrons. In antimony, $\ell_{ph}$ is non-monotonous, but in WTe$_2$, it  shows only a mild shoulder. This indicates that normal ph-ph scattering in WTe$_2$ is weak and there is no clear regime of phonon hydrodynamics. (b) The phonon specific heat divided by the cube of temperature in the two semimetals. It is flat below 5 K and smoothly decreases with warming, indicating that phonon density of states (PDOS) is quadratic and anharmonicity is weak, in contrast to Sb \cite{ref10}, where PDOS has a non-trivial frequency dependence. (c) The amplitude of the exponent of the electric resistivity ($\rho=\rho_0+T^\gamma$) in WTe$_2$ and in Sb \cite{ref10}. In both cases, the smooth increase of $\gamma$ with cooling is suddenly disrupted at a temperature below which Umklapp ph-ph scattering becomes impossible. In both cases, below $\approx 10$ K, only e-e (and not e-ph) collisions contribute to decay of momentum flow. }
		\label{fig4}
	\end{figure*}
	
	Figure \ref{fig4}a compares the phonon mean-free-path in WTe$_2$ and in Sb(See the supporting information for details on how to obtain the phonon mean-free-path and phonon specific heat). Despite the comparable order of magnitude, there is a qualitative difference. In Sb \cite{ref10}, the temperature dependence of the mean-free-path is non-monotonous as observed in black P \cite{ref14}, in graphite \cite{ref15} and in Bi \cite{ref37}. In all these cases, this was attributed to strong normal (that is non-Umklapp) scattering among phonons. In a limited temperature window, they lead to an increase in the mean-free-path with warming. This feature is absent in WTe$_2$. Interestingly, as seen in Figure \ref{fig4}b, the temperature dependence of the specific heat is remarkably different in the two systems. In WTe$_2$, the Debye approximation holds below 5 K where $C_{\rm{\textit{ph}}}\propto T^3$ and a downward deviation starts afterwards. In Sb, the specific heat has a non-trivial temperature dependence indicating strong anharmonicity.  Interestingly, the temperature dependence of specific heat in bismuth is also non-trivial \cite{ref41}. This correlation between two distinct experimental features confirms that anharmonicity (i.e. a non-quadratic phonon density of states) amplifies normal ph-ph collisions, indispensable for phonon hydrodynamics. 
	
	Scrutinizing the temperature dependence of resistivity reveals another important feature of the interplay between phonons and electrons. In the Bloch-Gr\"uneisen picture of resistivity driven by electron-phonon scattering, the electrical resistivity is linear at high temperature (when the electron-phonon scattering is quasi-elastic) and becomes $\propto T^5$ at low temperatures when the population of acoustic phonons and their typical wave-vector rapidly shrink. The exponent of electric resistivity (following $\rho=\rho_0+T^\gamma$) can be extracted using $\gamma=\frac{\partial {\rm{ln}}(\rho-\rho_0)}{\partial {\rm{ln}}T}$ \cite{ref10,ref42}. Figure \ref{fig4}c compares $\gamma$ in WTe$_2$ and in Sb. One can see that in both cases, the Bloch-Gr\"uneisen picture suddenly breaks down below 10 K. Resistivity becomes purely $\propto  T^2$. This becomes understandable by taking into account the fact that ph-ph collisions are no more Umklapp below 10 K.Therefore, a momentum taken by the phonon reservoir through an e-ph collision will  eventually return back to the electron reservoir through another ph-e collision.  As a consequence, e-ph collisions  do not degrade the momentum flow of electrons. Note that in contrast to Sb,  in WTe$_2$, the phonon scattering time is much longer than the e-e scattering time (see the supporting information).
	
	\section{Conclusions}
	In summary, we studied the electrical and thermal transport properties of WTe$_2$ at low temperatures, and found that the finite temperature deviation from the WF law is amplified with sample quality as expected in the hydrodynamic picture of electron transport. Phonons are strongly scattered by electrons,  but there is no signature of purely phononic hydrodynamics, presumably due to the weakness of normal ph-ph collisions\cite{ref43}. At cryogenic temperatures, there is substantial momentum exchange between the electronic and phononic reservoirs. Phonons cannot loose momentum alone, but electrons can, and this leads to the disruption of the Bloch-Gr\"uneisen picture and the emergence of a purely $T^2$ resistivity.

	\noindent
	\\
	*\verb|zengwei.zhu@hust.edu.cn|
	\\
	*\verb|kamran.behnia@espci.fr|
\providecommand{\noopsort}[1]{}\providecommand{\singleletter}[1]{#1}%

	\textbf{DATA AVAILABILITY}\\
	The data underlying this article will be shared on reasonable request to the corresponding author.\\
	
	\textbf{ACKNOWLEDGEMENTS}\\
	This work was supported by The National Key Research and Development Program of China (Grant No.2022YFA1403500), the National Science Foundation of China (Grant No. 12004123, 51861135104 and 11574097 ), and the Fundamental Research Funds for the Central Universities (Grant no. 2019kfyXMBZ071).  Kamran Behnia was supported by the Agence Nationale de la Recherche (ANR-19-CE30-0014-04). Xiaokang Li acknowledges the China National Postdoctoral Program for Innovative Talents (Grant No.BX20200143) and the China Postdoctoral Science Foundation (Grant No.2020M682386).\\
	
	\textbf{COMPETING INTERESTS}\\
	The authors declare no competing interests.\\
	
	\textbf{AUTHOR CONTRIBUTIONS}\\
	Zengwei Zhu and Kamran Behina conceived this work; Wei Xie grow the samples. With the help from Feng Yang, Liangcai Xu and Xiaokang Li, Wei Xie performed the electric and thermal transport measurements. Wei Xie, Zengwei Zhu, and Kamran Behina wrote the manuscript with input from all authors.\\

	\renewcommand{\thetable}{S\arabic{table}}
	\renewcommand{\thefigure}{S\arabic{figure}}
	\renewcommand{\thesection}{S\arabic{section}}
	\renewcommand{\theequation}{S\arabic{equation}}

	\setcounter{figure}{0}
	\setcounter{table}{0}
	\setcounter{section}{0}
	\setcounter{equation}{0}
	

	\clearpage
	
		\textbf{Supplementary information for "Purity-dependent Lorenz number, electron hydrodynamics and electron-phonon coupling in WTe$_2$"}
		
		\author{Wei Xie$^{1}$, Feng Yang$^{1}$, Liangcai Xu$^{1}$, Xiaokang Li$^{1}$, Zengwei Zhu$^{1,*}$ and Kamran Behnia$^{2,*}$}
		
		\affiliation{$^1$Wuhan National High Magnetic Field Center and School of Physics, Huazhong University of Science and Technology,  Wuhan,  430074, China\\
			$^2$Laboratoire de Physique et Etude des Mat\'{e}riaux (CNRS/UPMC),Ecole Sup\'{e}rieure de Physique et de Chimie Industrielles, Paris, 75005, France}
		
		
		\maketitle

		\section{Samples and Methods.}
		High quality single crystals of  WTe$_2$ were grown by Te-flux method. W(99.95\%) and excessive amounts of Te(99.999\%) powder were sealed in alumina ampoule, and then sealed in quartz tube. The quartz tube was heated up to 1050$^{\circ}$C over 12 h and kept for 12 h, then cooled down slowly to 700$^{\circ}$C. We obtained WTe$_2$ single crystals with different qualities by changing cooling rate. Shining crystals were mechanically separated from the flux with care.
		
		The transport measurements were performed with home-built one-heater-two-thermometers or one-heater-two-thermocouples setups, which allowed us to measure the electrical resistivity and the thermal conductivity with the same electrodes. Above 2 K, We performed them in a physical property measurement system (Quantum Design) and between 300 mK to 10 K, in a Leiden dilution refrigerator (CF-CS81-1700-Maglev). Between 300 mK and 4 K, CX-1010 Cernox chips and RuO$_2$ thermometers were used to measure the temperature, while the CX-1030 Cernox chips were used between 300 mK and 40 K. And the type E thermocouples were used between 20 K and room temperature. The thermal gradient in the sample was produced through a 10 k$\Omega$ chip resistor. The electrical transport measurements were measured by A standard four-probe method. 
		
		\section{Magnetoresistance and the electronic thermal conductivity}
\begin{figure*}[h]
			\centering
			\includegraphics[width=16cm]{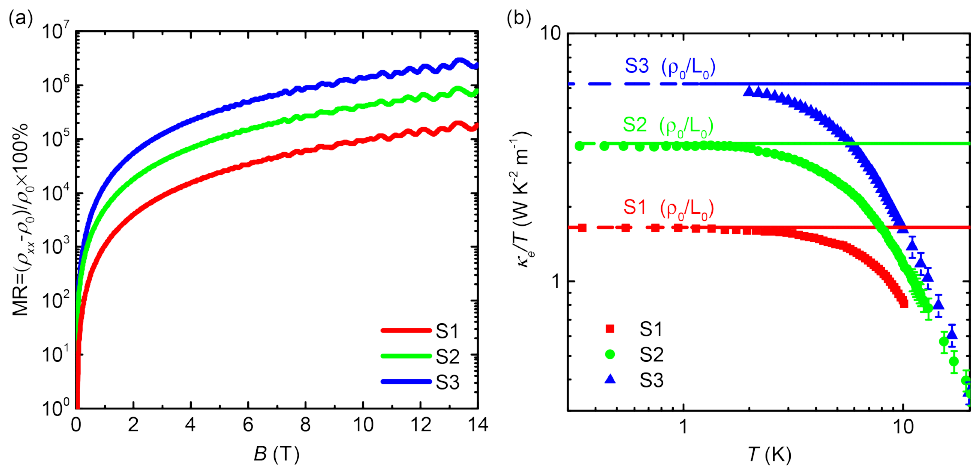}
			\caption{\textbf{Magnetoresistance and electronic thermal conductivity of WTe$_2$.} (a) Magnetoresistances of different samples at T = 2K. (b) Temperature dependence of the electronic thermal conductivity plotted as $\kappa_e$/T, the dashed lines represent $L_0/\rho_0$ for different samples.}
			\label{S1}
		\end{figure*}

		The large magnetoresistance of WTe$_2$ triggered the interest in this material \cite{refS1}. We observed a magnetoresistance as large as 10$^5\%$-10$^6\%$ at 2 K, when the applied field of 14 T, as shown in Figure \ref{S1}a. Magnetoresistance rapidly increases to 10$^4\%$-10$^5\%$ with a field of 3 T. We can separate phononic thermal conductivity by suppressing electronic thermal conductivity with a moderate magnetic field. Figure \ref{S1}b shows the electronic thermal conductivity plotted as $\kappa_e/T$, separated in this way. Dashed lines represent $L_0/\rho_0$ for different samples. The good agreement between them and $\kappa_e$/T at low temperature corresponds to verification of  WF law and verifies the accuracy of applying a magnetic field to separate the electronic and phononic thermal conductivity.
		
		\section{Specific heat and phononic mean free path.}
\begin{figure*}[h]
			\centering
			\includegraphics[width=7.5cm]{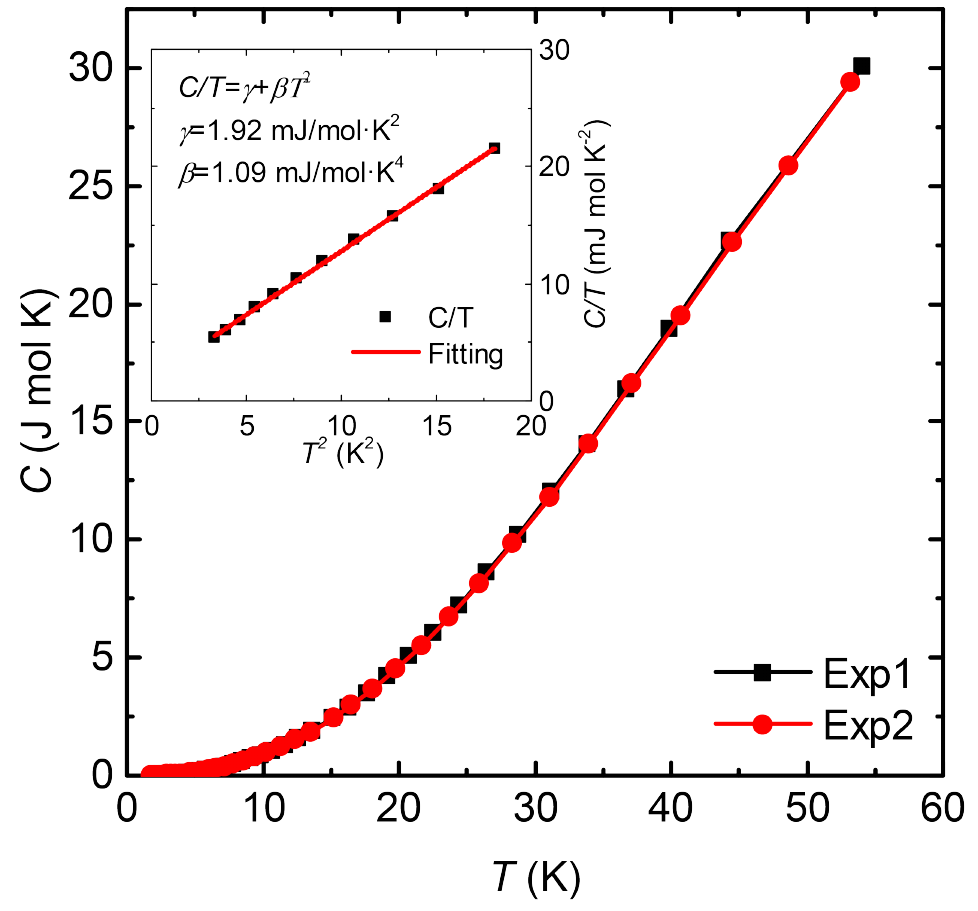}
			\caption{\textbf{Specific heat.} Temperature dependence of the specific heat in WTe$_2$. The red and black points represent two measurements of different samples. Inset shows a plot of\textit{ C/T} vs $T^2$ in the low temperature range. The red line corresponds to the fitting of $C/T=\gamma+\beta T^2$. }
			\label{S2}
		\end{figure*}

		Figure \ref{S2} shows the temperature dependence of specific heat. A  $C/T=\gamma+\beta T^2$ fit to the low temperature data is shown in the inset. The intercept corresponds to the electronic specific heat $\gamma$=1.92 mJ$\cdot$mol$^{-1}$$\cdot$K$^{-2}$ and the slope corresponds to the phononic specific heat $\beta$=1.09 mJ$\cdot$mol$^{-1}$$\cdot$K$^{-2}$. The phononic specific heat below 2 K is obtained by extrapolating the phononic specific heat measured at low temperature. This is reasonable, given the $T^3$ temperature dependence of phononic specific heat below 5 K. The phononic mean free path can be extracted using the relation $\ell_{ph}=\frac{3\kappa_{ph}}{C_{ph} \langle v_s \rangle}$, with $\langle v_s \rangle$=2200m$\cdot$s$^{-1}$\cite{refS2}. For different samples, $\ell_{ph}$ is shown in the main text.
		
		\section{ FFT analysis of the quantum oscillations.}
\begin{figure*}[h]
			\centering
			\includegraphics[width=16cm]{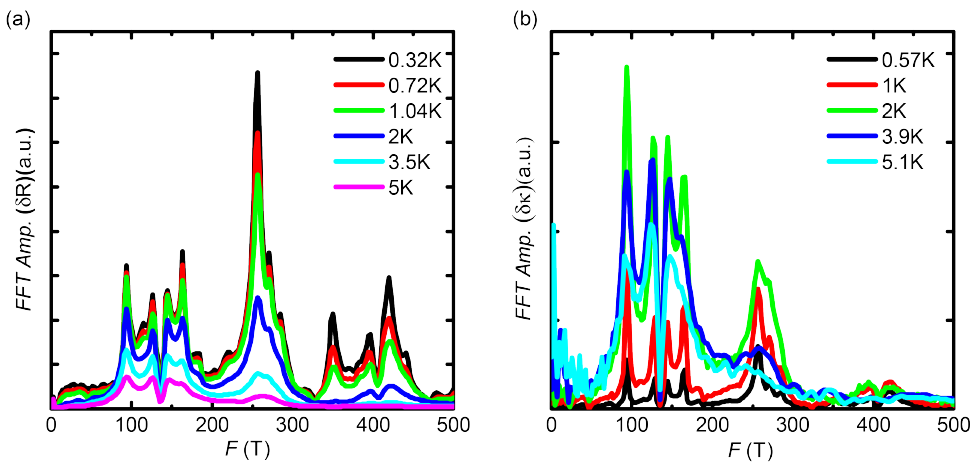}
			\caption{\textbf{The amplitude of the fast Fourier transform of $\delta$R and $\delta\kappa$ } (a) Fast Fourier transform analysis of the quantum oscillations of the magnetoresistance at several temperatures. (b) Fast Fourier transformation analysis of the quantum oscillations of the thermal conductivity at several temperatures. The four main frequencies can be observed at the quantum oscillations of magnetoresistance and thermal conductivity.}
			\label{S3}
		\end{figure*}

		Figure \ref{S3}a shows the results of the FFT of the oscillatory part of the magnetoresistance. Four main frequencies are observed.  The four frequencies are 93 T, 126 T, 144 T and 163 T, very close to the values reported before, and attributed to the two electron pockets and the two hole pockets \cite{refS3}. Quantum oscillation of thermal conductivity can be analyzed in the similar way to that of magnetoresistance. After subtracting a smooth background of the thermal conductivity, the oscillatory part is obtained. The FFT of the oscillatory part of the thermal conductivity is presented in Figure \ref{S3}b.  Four main frequencies consistent with those found in electrical conductivity are detected and shown in Figure \ref{S3}a.
		
		\section{ Dingle analysis.}
\begin{figure*}[h]
			\centering
			\includegraphics[width=17cm]{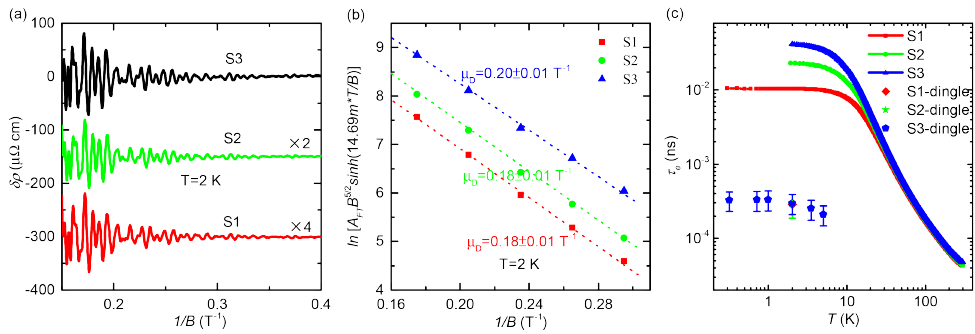}
			\caption{\textbf{Dingle analysis of WTe$_2$.} (a) Quantum oscillations of the magnetoresistance in three samples at T=2K. Curves are shifted vertically for clarity and multiplied by a factor 2 for S2 and 4 for S1. (b) Dingle analysis of the amplitude of the 93 T peak of the fast Fourier transform of $\delta\rho$ in different samples. The three dashed lines are almost parallel, indicating that the three samples have the same Dingle mobility. (c) Temperature dependence of the electronic scattering time extracted from the electrical resistivity and the Dingle scattering time extracted from the Dingle analysis for three WTe$_2$ samples}
			\label{S4}
		\end{figure*}

		Using $\mu_{tr}=1/(n+p)e\rho_0$, we can get the transport mobility $\mu_{tr}$=63 T$^{-1}$ of sample S3. As shown in Figure \ref{S4}, the Dingle analysis gives the Dingle mobility $\mu_{D}$=0.2 T$^{-1}$ of the same sample. The dashed line in Figure \ref{S4}b are almost parallel, indicating that the three samples have similar Dingle mobility, although the residual resistivity and the transport mobility of sample S1 and S3 is four times different, and the Dingle mobility of sample S3 is only about 1.1 times larger than that of sample S1, see the table \ref{Table3}. And $\mu_{tr}$ is more than 2 orders of magnitude larger than $\mu_{D}$, and the similar phenomenon has been reported in some semimetals before, such as Cd$_3$As$_2$, WP$_2$ and Sb, Table \ref{Table2} shows the two mobilities and the ratio of various semimetals. 
		
		As we can see in Figure \ref{S4}c, the electronic scattering time ($\tau_e$) extracted from the electrical resistivity is more than 2 orders of magnitude larger than the Dingle scattering time ($\tau_D$) extracted from the Dingle analysis in our cleanest sample S3, while in Cd$_3$As$_2$\cite{refS4}, WP$_2$\cite{refS5} and Sb\cite{refS6} with larger RRR, $\tau_e$ is 3-4 orders of magnitude larger than $\tau_D$.

		\section{The phonon thermal conductivity.}
\begin{figure*}[h]
			\centering
			\includegraphics[width=17cm]{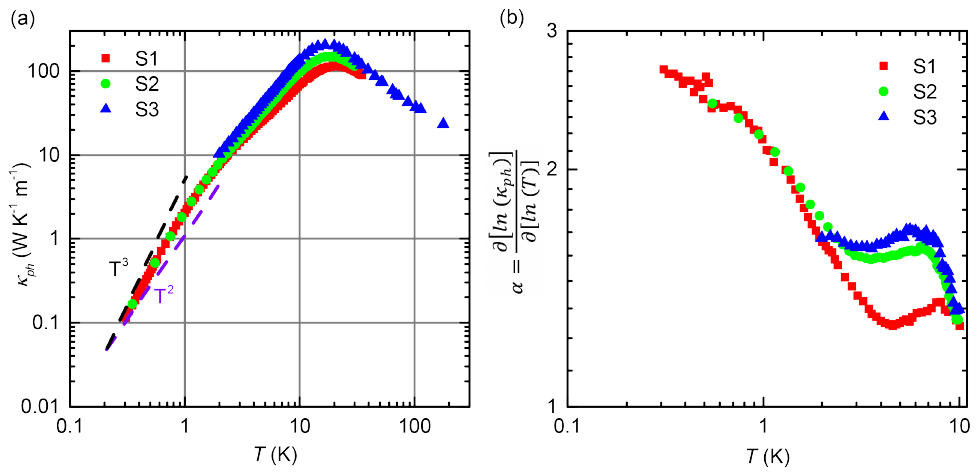}
			\caption{\textbf{(a) The temperature dependence of phonon thermal conductivity of three samples down to 0.3 K. (b) The evolution of the exponent of $\kappa_{ph}$ as a function of temperature below 10 K.}} 
			\label{S5}
		\end{figure*}

		Figure \ref{S5}a shows the temperature dependence of the phonon thermal conductivity of three WTe$_2$ samples down to 0.3 K. As we can see, at the lowest temperature, phonon thermal conductivity has an approximate cubic temperature dependence. The phonon thermal conductivity could display a $T^2$ thermal conductivity because of scattering by electrons. However,as seen in Figure \ref{S5}b, the exponent of $\kappa_{ph}$ smoothly evolves towards 3. While one cannot rule out that the temperature dependence of the phonon mean free path is affected by the presence of mobile electrons, there is no broad $T^2$ regime. Moreover, as one applies a magnetic field, neither the electron concentration, nor the electron mean-free-path changes. Therefore, it is unlikely that the phonon thermal conductivity is affected by magnetic field.

		\section{ $\kappa_e$ and $\kappa_{ph}$ in the two materials.}
\begin{figure*}[h]
			\centering
			\includegraphics[width=17cm]{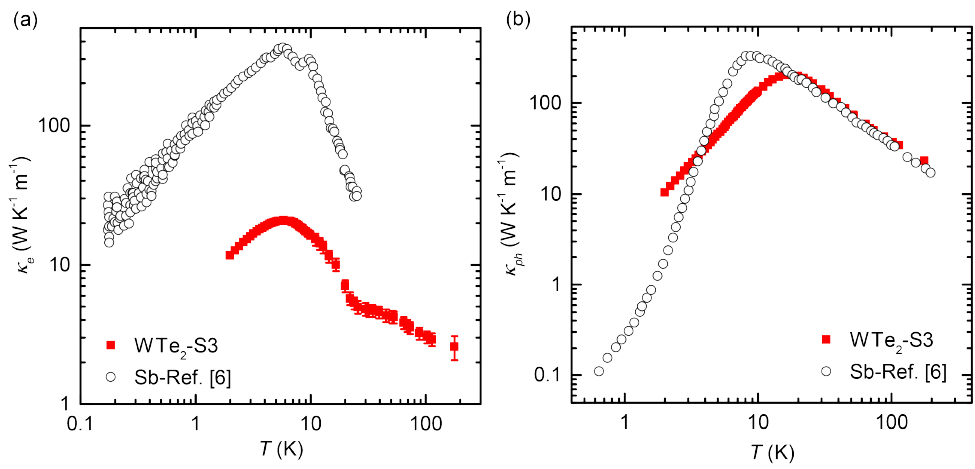}
			\caption{\textbf{Comparison of electronic and phononic thermal conductivity in WTe$_2$ and Sb.} (a) Temperature dependence of $\kappa_e$ and (b)$\kappa_{ph}$ in WTe$_2$ and Sb\cite{refS6}.  }
			\label{S6}
		\end{figure*}

		Figure \ref{S6}a shows $\kappa_e$ as a function of temperature in WTe$_2$ and Sb\cite{refS6}. One can see that $\kappa_e$ in WTe$_2$ is one order of magnitude smaller than that in Sb.  Figure \ref{S6}b shows the temperature dependence of $\kappa_{ph}$
		in the two materials. Above its maximum, $\kappa_{ph}$  is similar in the two systems. Below its peak, it decreases more rapidly in Sb, consistent with a stronger e-ph coupling in the latter compared to WTe$_2$.

		\begin{table*}[h]
			\centering
			\begin{tabular}{|c|c|c|c|c|c|c|c|c|}
				\hline
				Sample & Size(mm$^3$)(width$\times$thickness$\times$length) & RRR & $\overline{s}$(mm) & $l_0(um)$ & $\rho_0(u\Omega.cm)$ & A($n\Omega.cm.K^{-2}$) & B($n\Omega.cm.K^{-2}$) & B/A \\ \hline
				1 & 0.537×0.162×1.12 & 250 & 0.295 & 3.2 & 1.48 & 4.49±0.20 & 10.28±0.11 & 2.29 \\ \hline
				2 & 0.069×0.88×1.42 & 540 & 0.246 & 7.3 & 0.65 & 5.10±0.12 & 12.50±0.10 & 2.45 \\ \hline
				3 & 0.469×0.238×1.24 & 840 & 0.334 & 13.2 & 0.36 & 3.92±0.13 & 10.00±0.09 & 2.55 \\ \hline
			\end{tabular}
			\caption{\textbf{Details of the measured samples.} RRR is the residual resistivity ratio defined as $\rho_{300K}/\rho_{2K}$. $\overline{s}$ = $\sqrt{witdth\times thickness}$ represents the average diameter of the conducting cross-section. The carrier mean free path $l_0$ was calculated from the residual resistivity $\rho_0$ and the expression for Drude conductivity assuming four spherical hole and four spherical electron pockets. A and B are the electrical and thermal T$^2$-resistivities prefactors,respectively.}
			\label{Table1}
		\end{table*}
		\begin{table*}[h]
			\centering
			\begin{tabular}{|c|c|c|c|c|c|c|c|}\hline
				Sample & Carrier density ($10^{19}$ cm$^{-3}$) & RRR & $\rho_0$ ($\mu\Omega$.cm) & $\mu_{tr} (m^2V^{-1}s^{-1})$ & $\mu_D (m^2V^{-1}s^{-1})$ & r & Reference \\ \hline
				WTe$_2$ S1 & n=p=1.37 & 250 & 1.48 & 15 & 0.18 & 83 & This work \ \\ \hline
				WTe$_2$ S2 & n=p=1.37 & 540 & 0.65 & 35 & 0.18 & 194 & This work \\ \hline
				WTe$_2$ S3 & n=p=1.37 & 840 & 0.36 & 63 & 0.20 & 315 & This work \\ \hline
			\end{tabular}
			\caption{\textbf{The transport and quantum mobilities in the WTe$_2$ samples.} The transport mobility $u_{tr}$ is calculated from the residual resistivity $\rho_0$ and  carrier densities by 1/$\rho_0e(n+p)$. $u_D$ is the Dingle mobility extracted from a Dingle analysis of the quantum oscillations. r is the ratio of $\mu_{tr}$ to $\mu_D$. $\mu_D$ keeps almost constant while the transport mobility varies 4 times. This is similar to the case of Sb\cite{refS7}.}
			\label{Table3}
		\end{table*}

		\begin{table*}[h]
			\centering
			\begin{tabular}{|c|c|c|c|c|c|c|c|}\hline
				Sample & Carrier density ($10^{19}$ cm$^{-3}$) & RRR & $\rho_0$ ($\mu\Omega$.cm) & $\mu_{tr} (m^2V^{-1}s^{-1})$ & $\mu_D (m^2V^{-1}s^{-1})$ & r & Reference \\ \hline
				WTe$_2$ S3 & n=p=1.37 & 840 & 0.36 & 63 & 0.20 & 315 & This work \\ \hline
				Sb S1 & n=p=5.5 & 260 & 0.159 & 71 & 0.33 & 215 & \cite{refS7} \\ \hline
				Sb S2 & n=p=5.5 & 430 & 0.0946 & 120 & 0.36 & 333 & \cite{refS7} \\ \hline
				Sb S3 & n=p=5.5 & 3000 & 0.0134 & 848 & 0.38 & 2231 & \cite{refS7} \\ \hline
				Sb S4 & n=p=5.5 & 1700 & 0.0241 & 772 & 0.38 & 2031 & \cite{refS7} \\ \hline
				Cd$_3$As$_2$ & n=0.74 & 4100 & 0.021 & 870 & 0.087 & 10000 &  \cite{refS4} \\ \hline
				WP$_2$ & n=p=250 & 24850 & 0.005 & 400 & 0.08 & 5000 & \cite{refS5} \\ \hline
				NbAs & n=p=1 & 72 & — & 193 & 0.193 & 1000 & \cite{refS8} \\ \hline
				NbAs$_2$ & n=p=12 & 1580 & 0.041 & 62.5 & $\sim$0.1 & 625 & \cite{refS9} \\ \hline
				PtBi$_2$ & n=p=10 & 1667 & 0.024 & 6.55 & 0.0376 & 174 & \cite{refS10} \\ \hline
				LaBi & n=p=20 & 610 & 0.1 & 15.6 & 0.165 & 95 & \cite{refS11} \\ \hline
				TaAs & n=p=0.5 & 49 & 2 & 48 & 0.61 & 79 & \cite{refS12} \\ \hline
				NbP & n=p=0.15 & 115 & 0.63 & 500 & >10 & <50 & \cite{refS13} \\ \hline
				LaSb & n=p=16 & 170 & 0.6 & 3.3 & 0.125 & 26 & \cite{refS11} \\ \hline
				W$_2$As$_3$ & n=p=2 & 1240 & 0.23 & 3 & 0.13 & 23 & \cite{refS14} \\ \hline
				PrAlSi & n=p=4.75 & 4 & 15 & 0.5 & 0.33 & 1.5 & \cite{refS15} \\ \hline
				TaP & n=p=2 & 8 & 3 & 3.5 & 3.2 & 1.1 & \cite{refS16} \\ \hline
			\end{tabular}
			\caption{\textbf{The transport and quantum mobilities in different semimetals.} The transport mobility $u_{tr}$ is calculated from the residual resistivity $\rho_0$ and  carrier densities by 1/$\rho_0e(n+p)$. $\mu_D$ is the Dingle mobility extracted from a Dingle analysis of the quantum oscillations. r is the ratio of $\mu_{tr}$ to $\mu_D$. Some semimetal data extracted from the references are also listed in the table for comparison. }
			\label{Table2}
		\end{table*}

		%

\end{document}